%



\font\twelverm=cmr10 scaled 1200    \font\twelvei=cmmi10 scaled 1200
\font\twelvesy=cmsy10 scaled 1200   \font\twelveex=cmex10 scaled 1200
\font\twelvebf=cmbx10 scaled 1200   \font\twelvesl=cmsl10 scaled 1200
\font\twelvett=cmtt10 scaled 1200   \font\twelveit=cmti10 scaled 1200
\font\twelvesc=cmcsc10 scaled 1200  
\skewchar\twelvei='177   \skewchar\twelvesy='60


\def\twelvepoint{\normalbaselineskip=12.4pt plus 0.1pt minus 0.1pt
  \abovedisplayskip 12.4pt plus 3pt minus 9pt
  \belowdisplayskip 12.4pt plus 3pt minus 9pt
  \abovedisplayshortskip 0pt plus 3pt
  \belowdisplayshortskip 7.2pt plus 3pt minus 4pt
  \smallskipamount=3.6pt plus1.2pt minus1.2pt
  \medskipamount=7.2pt plus2.4pt minus2.4pt
  \bigskipamount=14.4pt plus4.8pt minus4.8pt
  \def\rm{\fam0\twelverm}          \def\it{\fam\itfam\twelveit}%
  \def\sl{\fam\slfam\twelvesl}     \def\bf{\fam\bffam\twelvebf}%
  \def\mit{\fam 1}                 \def\cal{\fam 2}%
  \def\sc{\twelvesc}               \def\tt{\twelvett}
  \def\sf{\twelvesf}
  \textfont0=\twelverm   \scriptfont0=\tenrm   \scriptscriptfont0=\sevenrm
  \textfont1=\twelvei    \scriptfont1=\teni    \scriptscriptfont1=\seveni
  \textfont2=\twelvesy   \scriptfont2=\tensy   \scriptscriptfont2=\sevensy
  \textfont3=\twelveex   \scriptfont3=\twelveex  \scriptscriptfont3=\twelveex
  \textfont\itfam=\twelveit
  \textfont\slfam=\twelvesl
  \textfont\bffam=\twelvebf \scriptfont\bffam=\tenbf
  \scriptscriptfont\bffam=\sevenbf
  \normalbaselines\rm}



\def\beginlinemode{\endmode
  \begingroup\parskip=0pt \obeylines\def\\{\par}\def\endmode{\par\endgroup}}
\def\beginparmode{\endmode
  \begingroup \def\endmode{\par\endgroup}}
\let\endmode=\par
{\obeylines\gdef\
{}}
\def\singlespace{\baselineskip=\normalbaselineskip}

\def\oneandahalfspace{\baselineskip=\normalbaselineskip
  \multiply\baselineskip by 3 \divide\baselineskip by 2}
\def\doublespace{\baselineskip=\normalbaselineskip \multiply\baselineskip by 2}

\newcount\firstpageno
\firstpageno=2
\footline={\ifnum\pageno<\firstpageno{\hfil}\else{\hfil\twelverm\folio\hfil}\fi}
\def\toppageno{\global\footline={\hfil}\global\headline
  ={\ifnum\pageno<\firstpageno{\hfil}\else{\hfil\twelverm\folio\hfil}\fi}}
\let\rawfootnote=\footnote              
\def\footnote#1#2{{\rm\singlespace\parindent=0pt\parskip=0pt
  \rawfootnote{#1}{#2\hfill\vrule height 0pt depth 6pt width 0pt}}}
\def\raggedcenter{\leftskip=4em plus 12em \rightskip=\leftskip
  \parindent=0pt \parfillskip=0pt \spaceskip=.3333em \xspaceskip=.5em
  \pretolerance=9999 \tolerance=9999
  \hyphenpenalty=9999 \exhyphenpenalty=9999 }
\def\dateline{\rightline{\ifcase\month\or
  January\or February\or March\or April\or May\or June\or
  July\or August\or September\or October\or November\or December\fi
  \space\number\year}}
\def\received{\vskip 3pt plus 0.2fill
 \centerline{\sl (Received\space\ifcase\month\or
  January\or February\or March\or April\or May\or June\or
  July\or August\or September\or October\or November\or December\fi
  \qquad, \number\year)}}


\hsize=6.5truein
\vsize=8.5truein  
\voffset=0.0truein
\parskip=\medskipamount
\def\\{\cr}
\twelvepoint            
\doublespace            
\overfullrule=0pt       




\def\title                      
  {\null\vskip 3pt plus 0.2fill
   \beginlinemode \doublespace \raggedcenter \bf}

\def\author                     
  {\vskip 3pt plus 0.2fill \beginlinemode
   \singlespace \raggedcenter\sc}

\def\affil                      
  {\vskip 3pt plus 0.1fill \beginlinemode
   \oneandahalfspace \raggedcenter \sl}

\def\abstract                   
  {\vskip 3pt plus 0.3fill \beginparmode
   \singlespace ABSTRACT: }

\def\endtopmatter               
  {\endpage                     
   \body}

\def\body                       
  {\beginparmode}               

\def\head#1{                    
  \goodbreak\vskip 0.5truein    
  {\immediate\write16{#1}
   \raggedcenter \uppercase{#1}\par}
   \nobreak\vskip 0.25truein\nobreak}

\def\subhead#1{                 
  \vskip 0.25truein             
  {\raggedcenter {#1} \par}
   \nobreak\vskip 0.25truein\nobreak}

\def\beginitems{
\par\medskip\bgroup\def\i##1 {\item{##1}}\def\ii##1 {\itemitem{##1}}
\leftskip=36pt\parskip=0pt}
\def\enditems{\par\egroup}

\def\beneathrel#1\under#2{\mathrel{\mathop{#2}\limits_{#1}}}

\def\refto#1{$^{#1}$}           

\def\references                 
  {\head{{\bf References}}            
   \beginparmode
   \frenchspacing \parindent=0pt \leftskip=1truecm
   \parskip=8pt plus 3pt \everypar{\hangindent=\parindent}}

\gdef\refis#1{\item{#1.\ }}                     

\gdef\journal#1, #2, #3, 1#4#5#6{               
    {\sl #1~}{\bf #2}, #3 (1#4#5#6)}            

\gdef\refa#1, #2, #3, #4, 1#5#6#7.{\noindent#1, #2 {\bf #3}, #4 (1#5#6#7).\rm}

\gdef\refb#1, #2, #3, #4, 1#5#6#7.{\noindent#1 (1#5#6#7), #2 {\bf #3}, #4.\rm}

\def\pr{\journal Phys.Rev., }

\def\prl{\journal Phys.Rev.Lett., }

\def\jmp{\journal J.Math.Phys., }

\def\np{\journal Nucl.Phys., }

\def\annp{\journal Ann.Phys.(N.Y.), }

\def\cqg{\journal Class.Quantum Grav., }

\def\endreferences{\body}

\def\figurecaptions             
  {\endpage
   \beginparmode
   \head{Figure Captions}
}

\def\endpage                    
  {\vfill\eject}

\def\endpaper                   
  {\endmode\vfill\supereject}


\def\heading                            
  {\vskip 0.5truein plus 0.1truein      
   \beginparmode \def\\{\par} \parskip=0pt \singlespace \raggedcenter}

\def\subheading                         
  {\vskip 0.25truein plus 0.1truein     
   \beginlinemode \singlespace \parskip=0pt \def\\{\par}\raggedcenter}

\def\tag#1$${\eqno(#1)$$}

\def\align#1$${\eqalign{#1}$$}

\def\aligntag#1$${\gdef\tag##1\\{&(##1)\cr}\eqalignno{#1\\}$$
  \gdef\tag##1$${\eqno(##1)$$}}

\def\overset #1\to#2{{\mathop{#2}\limits^{#1}}}
\def\underset#1\to#2{{\let\next=#1\mathpalette\undersetpalette#2}}
\def\undersetpalette#1#2{\vtop{\baselineskip0pt
\ialign{$\mathsurround=0pt #1\hfil##\hfil$\crcr#2\crcr\next\crcr}}}


\def\ref#1{Ref.~#1}                     
\def\Ref#1{Ref.~#1}                     
\def\[#1]{[\cite{#1}]}
\def\cite#1{{#1}}
\def\(#1){(\call{#1})}
\def\call#1{{#1}}
\def\taghead#1{}
\def\frac#1#2{{#1 \over #2}}
\def\half{{\frac 12}}

\def\12{{1\over2}}

\def\sla{\raise.15ex\hbox{$/$}\kern-.57em}
\def\leaderfill{\leaders\hbox to 1em{\hss.\hss}\hfill}
\def\twiddle{\lower.9ex\rlap{$\kern-.1em\scriptstyle\sim$}}
\def\bigtwiddle{\lower1.ex\rlap{$\sim$}}
\def\gtwid{\mathrel{\raise.3ex\hbox{$>$\kern-.75em\lower1ex\hbox{$\sim$}}}}
\def\ltwid{\mathrel{\raise.3ex\hbox{$<$\kern-.75em\lower1ex\hbox{$\sim$}}}}
\def\square{\kern1pt\vbox{\hrule height 1.2pt\hbox{\vrule width 1.2pt\hskip 3pt
   \vbox{\vskip 6pt}\hskip 3pt\vrule width 0.6pt}\hrule height 0.6pt}\kern1pt}
\def\tdot#1{\mathord{\mathop{#1}\limits^{\kern2pt\ldots}}}

\def\pmb#1{\setbox0=\hbox{#1}%
  \kern-.025em\copy0\kern-\wd0
  \kern  .05em\copy0\kern-\wd0
  \kern-.025em\raise.0433em\box0 }

\def\pp{{\prime\prime}}

\def\e{{\epsilon}}

\def\half{{1 \over 2}}
\def\ra{{\rangle}}
\def\la{{\langle}}

\def\l{{\lambda}}

\def\p{{\bf p}}

\def\x{{\bf x}}
\def\y{{\bf y}}

\def\ria{{\rightarrow}}

\catcode`@=11
\newcount\r@fcount \r@fcount=0
\newcount\r@fcurr
\immediate\newwrite\reffile
\newif\ifr@ffile\r@ffilefalse
\def\w@rnwrite#1{\ifr@ffile\immediate\write\reffile{#1}\fi\message{#1}}

\def\writer@f#1>>{}
\def\referencefile{
  \r@ffiletrue\immediate\openout\reffile=\jobname.ref%
  \def\writer@f##1>>{\ifr@ffile\immediate\write\reffile%
    {\noexpand\refis{##1} = \csname r@fnum##1\endcsname = %
     \expandafter\expandafter\expandafter\strip@t\expandafter%
     \meaning\csname r@ftext\csname r@fnum##1\endcsname\endcsname}\fi}%
  \def\strip@t##1>>{}}

\def\citeall#1{\xdef#1##1{#1{\noexpand\cite{##1}}}}
\def\cite#1{\each@rg\citer@nge{#1}} 

\def\each@rg#1#2{{\let\thecsname=#1\expandafter\first@rg#2,\end,}}
\def\first@rg#1,{\thecsname{#1}\apply@rg}   
\def\apply@rg#1,{\ifx\end#1\let\next=\relax
\else,\thecsname{#1}\let\next=\apply@rg\fi\next}

\def\citer@nge#1{\citedor@nge#1-\end-}  
\def\citer@ngeat#1\end-{#1}
\def\citedor@nge#1-#2-{\ifx\end#2\r@featspace#1 
  \else\citel@@p{#1}{#2}\citer@ngeat\fi}    
\def\citel@@p#1#2{\ifnum#1>#2{\errmessage{Reference range #1-#2\space is bad.}%
    \errhelp{If you cite a series of references by the notation M-N, then M and
    N must be integers, and N must be greater than or equal to M.}}\else%
 {\count0=#1\count1=#2\advance\count1 by1\relax\expandafter\r@fcite\the\count0,
  \loop\advance\count0 by1\relax
    \ifnum\count0<\count1,\expandafter\r@fcite\the\count0,%
  \repeat}\fi}

\def\r@featspace#1#2 {\r@fcite#1#2,}    
\def\r@fcite#1,{\ifuncit@d{#1}
    \newr@f{#1}%
    \expandafter\gdef\csname r@ftext\number\r@fcount\endcsname%
                     {\message{Reference #1 to be supplied.}%
                      \writer@f#1>>#1 to be supplied.\par}%
 \fi%
 \csname r@fnum#1\endcsname}
\def\ifuncit@d#1{\expandafter\ifx\csname r@fnum#1\endcsname\relax}%
\def\newr@f#1{\global\advance\r@fcount by1%
    \expandafter\xdef\csname r@fnum#1\endcsname{\number\r@fcount}}

\let\r@fis=\refis           
\def\refis#1#2#3\par{\ifuncit@d{#1}
   \newr@f{#1}%
   \w@rnwrite{Reference #1=\number\r@fcount\space is not cited up to now.}\fi%
  \expandafter\gdef\csname r@ftext\csname r@fnum#1\endcsname\endcsname%
  {\writer@f#1>>#2#3\par}}

\def\ignoreuncited{
   \def\refis##1##2##3\par{\ifuncit@d{##1}%
    \else\expandafter\gdef\csname r@ftext\csname r@fnum##1\endcsname\endcsname%
     {\writer@f##1>>##2##3\par}\fi}}

\def\r@ferr{\endreferences\errmessage{I was expecting to see
\noexpand\endreferences before now;  I have inserted it here.}}
\let\r@ferences=\references
\def\references{\r@ferences\def\endmode{\r@ferr\par\endgroup}}

\let\endr@ferences=\endreferences
\def\endreferences{\r@fcurr=0
  {\loop\ifnum\r@fcurr<\r@fcount
    \advance\r@fcurr by 1\relax\expandafter\r@fis\expandafter{\number\r@fcurr}%
    \csname r@ftext\number\r@fcurr\endcsname%
  \repeat}\gdef\r@ferr{}\endr@ferences}


\let\r@fend=\endpaper\gdef\endpaper{\ifr@ffile
\immediate\write16{Cross References written on []\jobname.REF.}\fi\r@fend}

\catcode`@=12

\citeall\refto      
\citeall\ref        %
\citeall\Ref        %

\def\p{\partial}
\def\la{\langle}
\def\ra{\rangle}
\def\ria{\rightarrow}
\def\x{{\bar x}}

\def\e{\epsilon}

\def\pp{{\prime\prime}}

\def\e{{\epsilon}}

\def\half{{1 \over 2}}
\def\ra{{\rangle}}
\def\la{{\langle}}

\def\l{{\lambda}}

\def\p{{\bf p}}

\def\x{{\bf x}}
\def\y{{\bf y}}

\def\ria{{\rightarrow}}

\def\pp{{\prime\prime}}

\centerline{\bf Trajectories for the Wave Function of the Universe}
\centerline{\bf from a Simple Detector Model}

\vskip 0.2in
\author J.J.Halliwell 

\vskip 0.2in
\affil Theory Group Blackett Laboratory Imperial
College London SW7 2BZ UK
\vskip 0.5in
\centerline {\rm PreprintImperial/TP/99-0/37. Revised version. November, 2000}

\abstract {Inspired by Mott's (1929) analysis of particle tracks
in a cloud chamber, we consider a simple model for quantum
cosmology which includes, in the total Hamiltonian, model
detectors registering whether or not the system, at any stage in
its entire history, passes through a series of regions in
configuration space.  We thus derive a variety of well-defined formulas for
the probabilities for trajectories associated with the solutions
to the Wheeler-DeWitt equation. The probability distribution is
peaked about classical trajectories in configuration space. The
``measured'' wave functions still satisfy the Wheeler-DeWitt equation,
except for small corrections due to the disturbance of the
measuring device. With modified boundary conditions,
the measurement amplitudes essentially agree with an
earlier result of Hartle derived on rather different grounds. In
the special case where the system is a collection of harmonic
oscillators, the interpretation of the results is aided by the
introduction of ``timeless'' coherent states -- eigenstates of the
Hamiltonian which are concentrated about entire classical
trajectories.}

\endtopmatter
\endpage

\head{\bf 1. Introduction}

The focus of attention in quantum cosmology is the Wheeler-DeWitt equation,
$$
H \Psi = 0
\eqno(1.1)
$$
Here, the wave function$ \Psi $ is a functional of the
gravitational and matter fields on a three-surface, and it
describes the quantum state of a closed cosmological model
[\cite{Har3}]. The most striking and conceptually problematic
aspect of this equation is that it does not involve time
explicitly, severely complicating efforts to extract predictions
from it [\cite{Ish,Kuc}]. Amongst the many attempts to understand
this feature, one is to claim that ``time'', and indeed entire
histories of the universe, are already contained amongst the
arguments of the wave function, hence no time label is required
[\cite{Bar3,BuIs}]. Whilst these claims seem to be true at some
level in simple models of quantum cosmology, it presents us with
the interesting challenge of reformulating standard quantum theory
without the explicit use of time, and then demonstrating the
emergence of time and of classical trajectories. Although the
Wheeler-DeWitt equation in the form (1.1) is unlikely to be the
last word in quantum gravity, it does seem likely that whatever
replaces it will still be of a timeless nature. The loop variables
programme of Ashtekar and others, for example, certainly preserves
this feature [\cite{Ash}]. It is therefore of interest to
investigate this feature in simple models.

Many attempts to use and make sense of Eq.(1.1) have been made.
These attempts focus on simple (minisuperspace) models,
in which one has an $N$-dimensional configuration space, $ {\cal C}$
with coordinates $\x$, and the Hamiltonian operator has the form
$$
H = - \half \nabla^2 + V (\x )
\eqno(1.2)
$$
The signature of the metric is typically hyperbolic
so the Wheeler-DeWitt equation is like a Klein-Gordon equation
in curved space with a spacetime dependent mass term.
Associated with it is a Klein-Gordon current
$$
J = i \left( \Psi \nabla \Psi^* - \Psi^* \nabla \Psi \right)
\eqno(1.3)
$$
Like the Klein-Gordon equation, however, this does not
produce a positive probability density
except in very special cases (namely when there is a Killing
vector associated with $H$). It also vanishes for real wave functions.
One might also consider the Schr\"odinger inner product,
$$
\la \Psi_1 | \Psi_2 \ra = \int_{\cal C} d^N \x \ \mu ( \x)
\Psi_1^* (\x) \Psi_2 (\x)
\eqno(1.4)
$$
where $\mu (\x) $ is an appropriate measure, but the norm
$ \la \Psi | \Psi \ra $ typically diverges. In practice,
most uses of the Wheeler-DeWitt equation rely on
something like the ``WKB interpretation'', in which
in the oscillatory regime the wave function is written
in the form $ \Psi = C e^{iS} $, where $S$ is a solution
to the Hamilton-Jacobi equation. It is argued that
this wave function corresponds to an ensemble of
classical trajectories satisfying the first integral
$ p = \nabla S $, with $|C|^2 $ giving a measure on the
ensemble. Although probably correct it is somewhat heuristic
and can only be used in the oscillatory regime.
(See Ref.[\cite{Hal1}], for example, for a discussion of
these issues).

Recent more successful work with the Wheeler-DeWitt or similar
equations involves the induced inner product (also known as Rieffel
induction or refined algebraic quantization) [\cite{Rie,Mar1}].
Here, one considers eigenvalues of the Wheeler-DeWitt
operator,
$$
H | \Psi_{Ek} \ra = E | \Psi_{E'k'} \ra
\eqno(1.5)
$$
where $k$ is the degeneracy label for each $E$.
The spectrum is typically continuous in $E$ in which case
the states are normalized via Eq.(1.4) according to
$$
\la \Psi_{E k} | \Psi_{E' k'} \ra =
\delta (E - E') \delta_{k k'}
\eqno(1.6)
$$
and one can now see why
$ \la \Psi | \Psi \ra $ diverges. The induced inner product
is then, loosely speaking, to drop the factor $\delta (E -E')$ as $E$ and
$E'$ are set to zero. This procedure can be defined rigorously and
induces an inner product on the zero energy eigenstates.
(This procedure is not of course necessary when the spectrum is discrete).

Related to these issues is the
prevalent idea that any operations performed on the wave
function in the computation of physically interesting probabilities
should commute with $H$ [\cite{DeW,Kuc,Rov3}]. Mathematically, this is to
respect the symmetry of the theory, reparametrization invariance,
expressed by the constraint equation (1.1). Physically, it is
connected with the fact that the universe is a genuinely closed
system, and all realistic measurements are carried out from the
inside, so cannot displace the system from a zero energy
eigenstate of $H$.

Given these preliminaries, turn now to the questions one would like
to ask of the wave function of the system in order to extract
useful cosmological predictions from it. We are interested
in the notion that the wave function corresponds in some way
to a set of trajectories. Let us therefore ask the question,

\item{} What is the probability that the system is found in a
series of regions in configuration space, $\Delta_1, \Delta_2,
\cdots \Delta_n $?

\noindent Note that the question is stated in such a way that does
not involve time. There is no requirement that the system enters
one of the regions at a particular ``time'', or that the regions
are entered in a particular order. We cannot ask this because the
Wheeler-DeWitt equation does not know about such an ordering
parameter. In the classical case the corresponding situation
consists of a statistical ensemble of classical trajectories with
the same fixed energy.
The trajectories are simply curves in
configuration space and it is straightforward to determine the
probability that
a given curve passes through a given region at any stage in its
entire history. The question is more involved in quantum theory,
since quantum theory is somewhat resistant to the notion of a
trajectory (in the non-relativistic case, it involves specifying
positions at different time, which do no commute). It is,
nevertheless, important to develop this notion, since the timeless
nature of the Wheeler-DeWitt equation cries out for an interpretation in
terms of entire histories of the universe.
The aim of this paper
is to offer one possible way of giving meaning to the above
question in the quantum theory of simple cosmological models.

Intuitively, one would expect that the question can be formulated
and answered using a simple toolbox of parts: the quantum state $
| \Psi \ra $ satisfying the constraint, projection operators
onto the regions $\Delta_k$, or maybe projections
onto some class of operators which commute with the constraint
$H$. One might also expect to find the Green function
associated with the Wheeler-DeWitt equation, which has the form,
$$
G(\x, \y) =  i \int_0^{\infty} d \tau \ \la \x | e^{ - i (H - i \e )
\tau } | \y \ra = \la \x | { 1 \over ( H - i \e ) } | \y \ra
\eqno(1.7)
$$
(this is the analogue of the Feynman propagator).
It might possibly also involve
one of the other types of propagators obtained by integrating
$\tau$ over an infinite range in this expression [\cite{Tei,Hal3,HaOr}].
The question is then exactly how
one stitches all these components together to make a plausible
probability distribution describing trajectories passing through
a series of regions.

The decoherent histories approach offers an approach to answering
this question and it does indeed use some of the above elements
[\cite{DH}]. It is particularly adapted to this sort of situation
since it directly addresses the issue of defining a quantum notion
of ``trajectory'' and this approach is currently being
investigated in this context [\cite{HaTh}] (see also
Ref.[\cite{Har3}]). Other approaches involving observables --
operators commuting with the constraint -- have also been
considered [\cite{Rov1,Rov2,Mar1,Mar2}]. Most importantly,
Kiefer and Zeh [\cite{Zeh}] and Barbour [\cite{Bar1,Bar2,Bar3}]
have devoted much effort to elucidating the emergence of
trajectories and of time from the timeless Wheeler-DeWitt equation.

The approach we adopt here stems from Barbour's observation [\cite{Bar2}]
that a
substantial insight into the Wheeler-DeWitt equation may be found
in Mott's 1929 analysis of alpha-particle tracks in a Wilson
cloud chamber [\cite{Mot}]. Mott's paper concerned the question of how the
alpha-particle's outgoing spherical wave state, $ e^{ikR}/R $,
could lead to straight line tracks in a cloud chamber. His
explanation was to model the cloud chamber as a collection of
atoms that may be ionized by the passage of the alpha-particle.
They therefore act as measuring devices that measure the
alpha-particle's trajectory. The probability that certain atoms
are ionized is indeed found to be strongly peaked when the atoms
lie along a straight line through the point of origin of the
alpha-particle.

Although Mott seems to have had in mind a time-evolving process,
he actually solved the time-independent equation
$$
\left( H_0 + H_d + \lambda H_{int} \right) | \Psi \ra = E | \Psi \ra
\eqno(1.8)
$$
Here $H_0$ is the alpha-particle Hamiltonian, $H_d$ is the
Hamiltonian for the ionizing atoms, and $H_{int}$ describes the
Coulomb interaction between the alpha-particle and the ionizing
atoms (where $\lambda$ is a small coupling constant). Now the
interesting point, as Barbour notes, is that Mott derived all the
physics from this equation with little reference to time. Mott's
calculation is therefore an excellent model for many aspects of
the Wheeler-DeWitt equation. Barbour has elucidated this very
eloquently, showing how it sheds light on a number of different
aspects [\cite{Bar2,Bar1,Bar3,Bro}].

Barbour's discussion is largely qualitative. The point of the
present paper, by contrast, is to extract quantitative
information from the comparison between the Mott calculation and
the Wheeler-DeWitt equation. Mott derived the straight line tracks
by looking at the wave function associated with two atoms being in
the ionized state, for the special case of an outgoing wave
initial state. But since this is elementary quantum mechanics, it is a
simple matter to generalize it to arbitrary initial states and
other types of detectors models, and to derive a detailed
expression for the probability distribution. Therefore, {\it
Mott's calculation points the way towards a general expression
for the probability distribution for the system passing through a
series of regions in configuration space without reference to
time}. This is what we will work out in detail in this paper.

We consider a system in an $N$ dimensional configuration space
$R^N$ with coordinates $\x$ described by a Hamiltonan $H_0$, which
may be of the form (1.2), but the simplest case we consider is
a free particle. It is coupled to a set of  detectors via an
interaction $H_{int}$ and the state of the whole system is given
by the solution to (1.8). Mott used the electronic degrees of
freedom of atoms as detectors. However, the essence of the calculation is
maintained with a much simpler detector model. The detector we use
consists of a two state system with $H_d = 0$ and detector states
$ |0 \ra $ and $ | 1 \ra $, where,
$$
a | 0 \ra = 0, \quad a | 1 \ra = | 0 \ra,
\quad a^{\dag} |0 \ra = | 1 \ra, \quad a^{\dag} | 1 \ra = 0
\eqno(1.9)
$$
We take
$$
H_{int} = \sum_k \ f_k (\x) \left(a_k + a_k^{\dag} \right)
\eqno(1.10)
$$
Here, $ f_k (\x) $ is spatially localized in the region
$\Delta_k$. One could, for
example, take $f_k $ to be a window function which is $1$ in
$\Delta_k$ and $0$ outside it, but we will not restrict to this
choice. If the detector in $\Delta_k$  is ``initially'' in the  ground
state $ | 0 \ra $, it will be displaced into the excited state $ |
1 \ra $ if the particle's trajectory $\x (t) $ enters $\Delta_k$
and stays in the ground state otherwise. Of course, in the
timeless context of the Wheeler-DeWitt equation, ``initially'' has
no meaning. Instead, following Mott, the appropriate condition to
impose is that the detector is in the ground state in the absence
of coupling to the system. This detector is far from realistic,
not least of all because it can return to its ground state if the
particle spends too much time in the detector region. We will
discuss its problems and possible improvements. We note however,
that similarly simple detector models have been profitably used elsewhere,
{\it e.g.}, in the Coleman-Hepp model [\cite{CoHe}]. (See also
Ref.[\cite{Hal4}]).

It is perhaps worth noting that
the question considered here bears a close resemblance to the
arrival time and tunneling time questions in non-relativistic
quantum mechanics [\cite{Time,YaT}]. There also, time enters
in a non-trivial way, and equivalent classical approaches
to the problem are inequivalent at the quantum level.
A variety of approaches have been brought to bear on these
problems, including, as here, explicit detector models.

In Section 2, we solve the system Eq.(1.8) using the simple two
state detector model.

Using the results, we then ask, in Section 3,
some simple questions of the detected wave function. Does the detected
wave function still obey the Wheeler-DeWitt equation? We find that
it does exactly outside the detector region, and that it does
approximately (for small $\lambda$) inside the detector region. We
compute the probabilities for detection and see, as Mott
essentially saw, that they are strongly peaked when the
detectors lie along a classical trajectory.
We also observe that the resulting amplitude for detection
bears a very close resemblance to a formula written
down by Hartle [\cite{Har2}] (without using an explicit
detector model) and we discuss the connections with his result.
We also discuss various other aspects of the solution
in relation to detection and the
timelessness of the solutions.

The results of Section 3 indicate that the wave function of the
system may, in some sense be regarded as a superposition of states
each of which is peaked about an entire classical history. To
demonstrate this explicitly, we specialize, in Section 4, to the
case of a collection of harmonic oscillators and introduce a new
type of coherent state, the ``timeless coherent states''. These
states are eigenstates of the Hamiltonian, and are therefore
time-independent, but are peaked about classical trajectories. Any
eigenstate may be expanded in terms of theses states, and we show
that a series of detections along a classical path essentially
projects the state down onto a timeless coherent state.

Since the detector is so simple, its dynamics may be solved
exactly and this is carried out in Section 5. This calculation
confirms that the detector model is only physical realistic
in the perturbative regime (when the particle spends only
a short time in each detector region). Although
the solution is exact and leaves the boundary conditions
general, it turns out that it is not very useful for
the Mott solution, since the boundary conditions lead to a
rather inelegant solution. On the other hand, it is by no
means clear that one is required to take the Mott boundary
conditions for the analagous situation in quantum cosmology,
and given the freedom to choose different conditions,
an elegant alternative solution for the detector amplitude
is obtained. It is in fact almost the same as the amplitude
Hartle proposed [\cite{Har2}].

In Section 6, we briefly describe a more elaborate detector
model, in which the detector is a simple harmonic oscillator
coupled to the particle with the same coupling (1.10).
The solution has a nice path integral representation
and suffers fewer shortcomings that the two-state detector.
It also clearly illustrate the peak about classical paths.

We summarize and conclude in Section 7.

\head{\bf 2. Detection Amplitude from the Two-State Detector}

We now solve the eigenvalue equation
$$
H | \Psi \ra = E | \Psi \ra
\eqno(2.1)
$$
with $H$ given by Eq.(1.8), and the detector is the simple two
state detector described in Section 1, with $H_d = 0 $. We will
for convenience refer to this equation as the Wheeler-DeWitt
equation (and it is convenient to retain a non-zero value of $E$).
It now reads
$$
(H -E)| \Psi \ra = (H_0 + \lambda H_{int} - E ) | \Psi \ra = 0
\eqno(2.2)
$$
We initially use only two detectors, so
$$
H_{int} = \sum_{k=1}^2 \ f_k (\x) (a_k + a_k^{\dag} )
\eqno(2.3)
$$
but the generalization to a arbitrary number of detectors
is straightforward.
We solve perturbatively by writing
$$
| \Psi \ra = | \Psi^{(0)} \ra + \l | \Psi^{(1)} \ra + \l^2 |
\Psi^{(2)} \ra + \cdots
\eqno(2.4)
$$
We require that in the absence of coupling to the detectors, the
detectors are in the state of no detection, $ | 0 \ra $. This means
that
$$
| \Psi^{(0)} \ra = | \psi \ra | 0 \ra | 0 \ra
\eqno(2.5)
$$
and $ | \psi \ra $ is the state we are trying to measure.
Inserting in (2.2) and equating powers of $\l$, we get,
$$
\eqalignno{
( H_0 - E ) | \Psi^{(0)} \ra &= 0
&(2.6) \cr
( H_0 - E ) | \Psi^{(1)} \ra &=  - H_{int} | \Psi^{(0)} \ra
&(2.7) \cr
( H_0 - E ) | \Psi^{(2)} \ra &=  - H_{int} | \Psi^{(1)} \ra
&(2.8) \cr}
$$
and similarly to higher orders. The first relation says that $| \psi \ra$
must obey the unperturbed eigenvalue equation, as expected.
Inserting (2.5) into Eq.(2.7), we get
$$
(H_0 - E ) | \Psi^{(1)} \ra =  - f_1 (\x ) | \psi \ra  | 1 \ra  | 0 \ra
- f_2 (\x ) | \psi \ra  | 0 \ra | 1 \ra
\eqno(2.9)
$$
This is readily solved by writing
$$
| \Psi^{(1)} \ra =
| \Psi^{(1)}_{00} \ra |0\ra |0\ra
+ | \Psi^{(1)}_{10} \ra |1\ra |0\ra
+ | \Psi^{(1)}_{01} \ra |0\ra |1\ra
+ | \Psi^{(1)}_{11} \ra |1\ra |1\ra
\eqno(2.10)
$$
and we discover that
$$
\eqalignno{
(H_0 - E) | \Psi^{(1)}_{00} \ra &= 0
&(2.11)\cr
(H_0 - E) | \Psi^{(1)}_{10} \ra &= - f_1 (\x) | \psi \ra
&(2.12)\cr
(H_0 - E) | \Psi^{(1)}_{01} \ra &= - f_2 (\x) | \psi \ra
&(2.13)\cr
(H_0 - E) | \Psi^{(1)}_{11} \ra &= 0
&(2.14)\cr }
$$
Eqs.(2.12), (2.13) may be solved with the assistance of the Green function $G$,
defined by Eq.(1.7) (with $H$ replaced by $H_0 - E $). It obeys the
equation
$$
( H_0 - E ) G = 1
\eqno(2.15)
$$
(For convenience we use an operator notation in which
$G$ is the operator with coordinate representation
$ G(\x, \y) = \la \x | G | \y \ra $ and the right-hand
side of (2.15) would be the delta-function $\delta (\x,\y)$ in
the coordinate representation).
We then obtain
$$
\eqalignno{
| \Psi^{(1)}_{10} \ra &= - G f_1 (\x) | \psi \ra + | \phi_1 \ra
&(2.16) \cr
| \Psi^{(1)}_{01} \ra &= - G f_2 (\x) | \psi \ra + | \phi_2 \ra
&(2.17) \cr}
$$
where $| \phi_{1,2} \ra $ are solutions to the homogeneous equation
$$
(H_0 - E) | \phi_{1,2} \ra = 0
\eqno(2.18)
$$
To fix some of these solutions more precisely, we need to appeal
to boundary conditions. This is a subtle issue and depends very much
on the precise context. Mott was concerned with the particular case
of an outgoing spherical wave and imposed boundary conditions
appropriate to this case. This led him to set $| \phi_{1,2} \ra$
and $ | \Psi_{11}^{(1)} \ra $ to zero (since otherwise it represents
a stream of incoming particles fired at an already excited detector)
[\cite{Mot}]. We are not obviously compelled to make the same
choice of boundary conditions in quantum cosmology, and we
will return to a discussion of this important issue in
Section 3(C). But for the moment, we work with the Mott solution.

At first order only one detector is stimulated into the excited
state. The system wave function correlated with detector
state $ | 1 \ra $ now is
$$
| \psi_1 \ra =  - \l G f_1 | \psi \ra
\eqno(2.19)
$$
and the probability that the detector registers is therefore
$$
p_1 = \la \psi_1  | \psi_1 \ra = \l^2 \la \psi | f_1 G^{\dag} G f_1 | \psi \ra
\eqno(2.20)
$$
(When the spectrum of $H_0$ is continuous, expressions of this type
need to be regularized along the lines of Eq.(1.6), but we will
carry out this explicitly only when we need to calculate it in
more detail).

To get two detectors to register, we need to go to second order.
We now have, from (2.8), and the solution (2.10),
$$
\eqalignno{
( H_0 - E ) | \Psi^{(2)} \ra &=
- \left( f_1 (a_1 + a_1^{\dag} ) + f_2 (a_2 + a_2^{\dag} ) \right)
| \Psi^{(1)} \ra
\cr
& =f_2 G f_1 | \psi \ra | 1 \ra | 1 \ra
+ f_1 G f_2 | \psi \ra | 1 \ra | 1 \ra + \cdots
&(2.21) \cr}
$$
where the omitted terms on the right-hand side are proportional
to $ | 0 \ra | 0 \ra $, $ | 0 \ra | 1 \ra $ and $ | 1 \ra | 0 \ra $,
and will not be needed. Again we may solve by expanding as
in (2.10). Here we write down only the term required, which
describes two detectors being excited
$$
| \Psi^{(2)} \ra = | \Psi^{(2)}_{11} \ra |1\ra |1\ra + \cdots
\eqno(2.22)
$$
and it is readily seen that the solution is
$$
| \Psi^{(2)}_{11} \ra = \left( G f_2 G f_1 + G f_1 G f_2 \right)
| \psi \ra
\eqno(2.23)
$$
Again, following the spirit of the Mott solution,
possible homogeneous solutions are set to zero.
We now have the system wave function
correlated with two detectors registering: it is
$$
| \psi_2 \ra = \l^2
 \left( G f_2 G f_1 + G f_1 G f_2 \right)
| \psi \ra
\eqno(2.24)
$$
and the probability is $ \la \psi_2 | \psi_2 \ra $.

The analysis is readily extended to an arbitrary number of
detectors, but it is easy to anticipate the result from (2.24):
for $n$ detectors, the amplitude is
$$
| \psi_n \ra = \l^n \left( G f_n G f_{n-1} \cdots G f_2 G f_1 \right)
| \psi \ra + \ {\rm symmetrizations}
\eqno(2.25)
$$
where ``symmetrizations'' means add all possible permutations
of $ 1,2,3, \cdots n $. It is clear that these terms are there
to ensure that there is no preference in the order in which
each of the detectors registers, reflecting the genuinely
timeless nature of the underlying dynamics. Eq.(2.25)
is the main result of this Section.

\head{\bf 3. Properties of the Solution}

We have shown that the wave function for the whole system
when there are, for example, two detectors, takes the form
$$
| \Psi \ra = | \psi_0 \ra | 0 \ra | 0 \ra
+ | \psi_1 \ra \left( | 1 \ra | 0 \ra + | 0 \ra | 1 \ra \right)
+ | \psi_2 \ra | 1 \ra | 1 \ra
\eqno(3.1)
$$
We can now ask various questions of the detected wave function
$ | \psi_2 \ra $, or more generally, Eq.(2.25).

\subhead{\bf 3(A). Does the Detected Wave Function obey the Wheeler-DeWitt
Equation?}

As stated in the Introduction, a prevalent idea in quantum gravity
is that all observables should commute with the total Hamiltonian
[\cite{DeW,Kuc,Rov3}]. Related to this is the notion that
``measurements'' of the wave function (whatever this may mean in
general) should not displace the system from its eigenstate of the
Hamiltonian. Given that we have presented here an explicit model
of detection, it is perhaps of interest to ask to what extent this
idea holds up.

We have, using (2.15), and taking the simple
case $n=2$,
$$
( H_0 - E ) | \psi_2 \ra =
\l^2 \left(  f_2 G f_1 +  f_1 G f_2 \right)
| \psi \ra
\eqno(3.2)
$$
In configuration space, the right-hand side zero, except in the
detector regions $\Delta_1$, $\Delta_2$, because the functions
$f_1$ and $f_2$ are localized there. In these regions, it is of
order $\l^2$, which we regard as small in comparison to the terms
on the left. Hence the measured wave function approximately obeys
the Wheeler-DeWitt equation. This is no surprise. In standard
quantum mechanics, a physically measured system does not obey the
Schr\"odinger equation but has corrections due to the measuring
device. That it obeys the Wheeler-DeWitt equation only
approximately is not in conflict with exact reparametrization
invariance, since the wave function for the entire system always
obeys the Wheeler-DeWitt equation exactly. On the other hand, one
wonders whether it might not be possible to devise a detection
scheme in which the detected amplitude obeys the Wheeler-DeWitt
equation exactly. For example, given the simple toolbox of parts
outlined in the Introduction (such as the Green function (1.7)),
it would not be unreasonable to guess that the detection amplitude
might be of the form (2.25) but with the $G$ given by Eq.(1.7)
replaced by the one obtained by integrating $ \tau $ over an
infinite range. This gives a solution to the Wheeler-DeWitt
equation, and as a consequence, the modified detection amplitude
would also obey the Wheeler-DeWitt equation exactly.

\subhead{\bf 3(B). Amplitudes for Classical Paths}

Of greater interest is the question of the configurations
about which the amplitude (2.25) (or the associated probability)
is peaked. The amplitude may be written in an explicit coordinate
representation as
$$
\eqalignno{
\la \x_f | \psi_n \ra = \lambda^n
& \int d^N \x_n \cdots d^N\x_2 d^N \x_1
\ G(\x_f, \x_n) f_n (\x_n)
\cr & \times G (\x_n , \x_{n-1} )
f_{n-1} (\x_{n-1} )
\ \cdots \ G(\x_2, \x_1 ) f_1  (\x_1 ) \ \psi (\x_1)
&(3.3) \cr}
$$
(plus symmetrizations). It has the form of approximate projections onto
the regions $\Delta_k$ (exact projections if the $f_k$ are window
functions) with evolution between regions described by the fixed energy
propagator $G(\x_{k+1}, \x_k) $. It is analogous to the amplitude
for a history of positions at a sequence of times in non-relativistic
quantum mechanics, which is known to be peaked about classical
trajectories [\cite{Hal2}]. But note that here the evolution is
at fixed values of the energy, and there is no reference to
time.

We can estimate the form of (3.3) using a WKB
approximation for the fixed energy propagator [\cite{Sch}]. It is given by an
expression of the form,
$$
G (\x^{\pp}, \x' ) = C  (\x^{\pp}, \x') e^{ i S (\x^{\pp}, \x') }
\eqno(3.4)
$$
where $C$ is a slowly varying prefactor and $S (\x^{\pp}, \x') $
is the fixed energy Hamilton-Jacobi function, i.e., the action
of the classical solution from $\x'$ to $\x^{\pp}$ with fixed
energy $E$ [\cite{Gol}]. The initial and final
momenta of this classical solution are
$$
\p^{\pp} = \nabla_{\x^\pp} S (\x^{\pp}, \x'),
\quad \p' = - \nabla_{\x'} S (\x^{\pp}, \x')
\eqno(3.5)
$$
In terms of (3.4), the amplitude (3.3) may be written,
$$
\eqalignno{
\la \x_f | \psi_n \ra = \lambda^n
\int d^N \x_n \cdots & d^N\x_2 d^N \x_1
\prod_{k=1}^n C (\x_{k+1}, \x_k) f_k (\x_k)
\cr \times &
\ \exp \left( i \sum_{k=1}^n S( \x_{k+1}, \x_k ) \right)
\psi (\x_1 )
&(3.6) \cr }
$$
where $\x_{n+1} = \x_f $.

Consider the integrals over $\x_2 \cdots \x_n $
with $\x_1$ (and $\x_{n+1}$) fixed. Suppose for the moment that the
functions $f_k$ are absent so the integrals are unrestricted.
By the stationary phase approximation, the
dominant contribution to the integral comes
from the values of $\x_2 \cdots \x_n$
for which the phase is stationary, i.e., for which
$$
\nabla_{\x_j} \sum_{k=1}^n S( \x_{k+1}, \x_k ) = 0
\eqno(3.7)
$$
for $j=2, \cdots n $. This equation means that
$$
\nabla_{\x_j} S (\x_{j+1}, \x_j ) + \nabla_{\x_j}
S( \x_j, \x_{j-1} ) = 0
\eqno(3.8)
$$
Using Eq.(3.5), this implies that the final momentum
of the classical path from $\x_{j-1}$ to $\x_j$ is equal
to the initial momentum of the classical path from
$\x_j$ to $\x_{j+1}$. It is not difficult to see that this
in turn implies that the point $\x_j$ must lie on the
classical path from $\x_{j-1}$ to $\x_{j+1}$. Hence,
the stationary phase points of the whole integral
(3.6) lie on the classical path from $\x_1$ to $\x_f$.
The approximate value of the integral is
the integrand of (3.6) with the stationary phase point values inserted.

Now consider what happens when the restricting functions $f_k$ are
present. If the regions $\Delta_k$ (about which the $f_k$ are
concentrated) include the stationary phase points (and if the
regions are larger than the fluctuations about these points),
then, since the integral takes it dominant contribution from these
points, the presence of the $f_k$ makes little difference and the
integral is given once again by its stationary phase value. On the
other hand, if one or more of the $f_k$ lie far away from the
stationary phase points, then, since the integral is prevented
from taking a contribution from these points, its value must be
much smaller than the stationary phase value. We thus see that the
amplitude (3.3) will be largest when the regions $\Delta_k$ are
chosen to include the stationary phase points of the integral. As
we have shown, the stationary phase points lie along the classical
path from $\x_1$ to $\x_f$. It follows that the amplitude (3.2)
will be largest when the regions $\Delta_k$ are chosen to lie
along a classical path.

Mott's argument for straightline paths in an expression
analogous to (2.25) relied on the explicit from of the
Green function in the free particle case, and on a special
initial state [\cite{Mot}].
Here we see that the peaking about classical
paths can be seen, at least heuristically, from elementary
properties of path integrals, for a broad class of Hamiltonians
and initial states.
Bell has also discussed the Mott calculation at some length
[\cite{Bel}]. He notes that the first projector $f_1$ spatially
localizes Mott's initial wave function, but in a realistic
ionizing event, the resultant uncertainty in momentum can still be
extremely small. As a consequence the angular spread of the wave
packet in its subsequent evolution can be extremely small, hence
the appearance of a straight line track.

\subhead{\bf 3(C). Comparison with a Result of Hartle}

The result (2.24) is very closely related to a result of
Hartle [\cite{Har2}]. He considered a simple model quantum cosmology
with a Hamiltonian quadratic in the momenta, as here, and asked
for the amplitude that the system passes through two regions
of configuration space, $\Delta_1$, $\Delta_2$. Using
some simple arguments about propagators
and elementary principles of quantum theory,
he showed that the amplitude is (in the language of the
present paper)
$$
\left( G f_2 G f_1 + G f_1 G f_2 \right)| \psi \ra
- \left( G^{\dag} f_2 G^{\dag} f_1 + G^{\dag} f_1 G^{\dag} f_2 \right)
| \psi \ra
\eqno(3.8)
$$
where $ \psi $ is a solution to the Wheeler-DeWitt equation
and $f_1$ and $f_2$ are taken to be exact projectors onto
the regions $\Delta_1$, $\Delta_2$. Other than the
factor of $\l^2$ in (2.24) which is
not important, Hartle's result differs
from (2.24) by the subtraction of an identical term but with
$G$ replaced by $G^{\dag}$ (which is generally not the same
as $G$).
Hartle argues that this should be
there on the grounds that, in an expression like (2.25)
with $G$ represented by (1.7),
the time parametrization should not have a preferred direction
with $\y$ the ``initial'' point of the parametrization
and $\x$ the ``final'' point, hence we should sum the amplitude
over both possible parametrization directions.

To understand why this term can be there in the present calculation,
let us first say a little more about the Green function.
$G$, as defined by Eq.(1.7), may be written more explicitly as
$$
G(\x, \y, E ) = \sum_n { u_n^* (\x) u_n (\y) \over E - E_n + i \e }
\eqno(3.9)
$$
where $u_n (\x) $ are eigenfunctions of $H_0$ with eigenvalue
$E_n$. $G (\x, \y, E ) $ is real when $E$ lies in the discrete
part of the spectrum and complex when $E$ lies in the continuous
part (see, for example, Ref.[\cite{PDX}]). Hence in the free
particle case considered by Mott, $G \ne G^{\dag}$, but for the
harmonic oscillator (considered in Section 4) $ G = G^{\dag}$.
Quantum cosmological models usually have a spectrum which is at
least in part continuous, so we expect $G \ne G^{\dag}$ in
general.

Now recall the comments after Eq.(2.18), where it was noted that
we are by no means obliged in quantum cosmology to take the same
boundary conditions as Mott. Since $G^{\dag}$ also satisfies
Eq.(2.15), we may use it in place of $G$ to generate solutions to
the detector amplitude. Because $ G - G^{\dag}  = i \delta (H_0 -
E) $, it is easily seen that the difference between using $G$ and
$G^{\dag}$ is a homogeneous solution in (2.16), (2.17). It is
also true that, in higher order perturbations, we may use $G$ or
$G^{\dag}$ or some combination. Therefore, there is a more general
class of solutions for the detection amplitude which are sums of
terms of the form (2.25) with some of the $G$'s replaced by
$G^{\dag}$'s (and with a suitable overall normalization). In
particular, Hartle's amplitude falls into this enlarged class of
solutions, so there is no conflict with his result.

Mott made a particular choice of solution appropriate to the
physical situation he as investigating. In the case of
relativistic field theory in Minkowski space, one would normally
impose some sort of causality requirement to fix the solution
more precisely, and so to choose between $G$ and $G^{\dag}$
(since $G$ is essentially the Feynman propagator). One could
require, for example, that the wave function for the whole system
is affected by the detector only in the future light cone of the
detector region. In quantum cosmology, however, although the
metric has hyperbolic signature like Minkowki space, it is by no
means clear that one is obliged to impose an analogous
requirement, and in fact it is difficult to see exactly how to do
this in general since the configuration space is usually not
globally hyperbolic. (See, however, Ref.[\cite{Tei}]). Instead,
one might expect to fix the choice of $G$ or $G^{\dag}$ by
appealing to cosmological boundary conditions. The no-boundary
proposal of Hartle and Hawking, for example, picks out a wave
function that is real [\cite{HaHa}]. It is sometimes claimed that
this corresponds to a ``time-symmetric'' wave function
[\cite{Haw}]. It therefore stays most closely to the timeless
nature of the Wheeler-DeWitt equation and in some sense represents
a complete abandonment of any fundamental notion of causality. It
is now interesting to note that Hartle's detection wave function
(3.8) is in fact real if $\psi$ is real. In many ways this
therefore seems like the most natural solution to take in the case
of the Wheeler-DeWitt equation. We do not, however, in this paper
commit to any particular choice of boundary condition.

In summary, therefore, a more general solution to the detector
model is a sum of terms of the form (2.25) involving both $G$ and
$G^{\dag}$, and this more general solution includes Hartle's
result (3.14). Note also that the replacement of $G$ by $G^{\dag}$
does not affect the discussion of the peaking of the amplitude
about classical trajectories.

\subhead{\bf 3(D). On Timelessness and Detection}

It is perhaps worth elaborating on a feature of the
detector model which appears at first sight to be incompatible
with the timelessness of the Wheeler-DeWitt equation. We
have coupled the system to a series of detectors via the
interaction Hamiltonian (1.8). This Hamiltonian describes a
situation in which, along a trajectory $ \x (t) $, the detector is
in the ground state ``before'' the trajectory enters the detector
region and in an excited state ``after'' it has passed through the
region. Along a {\it classical} trajectory $\x(t)$ in which there
is a notion of time, and of before and after, this is undeniably
correct. (The parameter $t$ simply labels the points along the
curve -- it may be take to be, for example, the distance along the
curve from some reference point.)
But how are we to understand how the detector works in
the genuinely timeless world described by the Wheeler-DeWitt
equation? There is no before and after and there is no preferred
direction of time.

The above results effectively imply that each solution to the
Wheeler-DeWitt equation may be regarded as a superposition of
states each of which is concentrated along an entire classical
trajectory in configuration space (and we will see this in more
detail in Section 4). What then seems to be happening in the
present model including the detector, is that each trajectory
carries a label indicating whether or not it passes through the
detector region at any stage along its entire length. In the
(perhaps restrictive) language of time, at each point along the
trajectory, the label allows the trajectory to ``know'' whether it
passed through the detector region in the past, or will pass
through it in the future. But it is only on adopting this temporal
language that the situation seems paradoxical. The paradox
vanishes when ones speaks the vocabulary of entire trajectories in
configuration space, and one can see this in the solution (3.1).
The wave function of the entire system is written as a correlated
state in which the state correlated with the detectors in the $ |
1 \ra $ state is the state (2.25): the detected state is indeed
concentrated on trajectories that pass through $\Delta_1, \cdots
\Delta_n$.

The issues discussed in this Section may be of relevance to the
perennial debate on the question of time asymmetry in quantum
cosmology [\cite{asy,Zeh}].


\head{\bf 4. Coherent States for Timeless Dynamics}

Since the peaking about classical paths is the most important
property of the amplitude (2.25), it is worth exploring it in more
detail for the special case of a collection of harmonic
oscillators, where it is possible to show very clearly how the
solutions to the Wheeler-DeWitt equation correspond to
superpositions of states peaked about classical paths. We will
introduce a class of coherent states appropriate to the timeless
theories considered here and which are natural analogues of the
standard coherent states of the harmonic oscillator.

The Hamiltonian for a set of $N$ identical harmonic oscillators is
$$
H_0 = \half \left( \p^2 + \x^2 \right)
\eqno(4.1)
$$
In this case the spectrum of $H_0$ is discrete and we have
$$
\delta (H_0 - E ) = \int_0^{2 \pi} { dt \over 2 \pi} e^{ - i (H_0 - E)t }
\eqno(4.2)
$$
Since $\delta (H_0 - E) $ is now in fact a true projection operator
we may write $ \delta (H_0 - E )^2 = \delta ( H_0 - E ) $ without
having to worry about regularization through the induced inner product,
as in the continuous case. (In this expression $E$ is allowed to take
only the discrete values corresponding to the spectrum of $H_0$).
The Green function $G$ is given as before by (1.7) with $H$
replaced by $H_0 - E $. For a one-dimensional oscillator, Eq.(4.2)
is equivalent to
$$
\delta (H_0 - E ) = | E \ra \la E |
\eqno(4.3)
$$
where $ | E \ra $ is the energy eigenstate. In more than one dimension
the energy eigenstates are degenerate, so Eq.(4.2) has the form
$$
\delta (H_0 - E ) = \sum_d \ | E, d \ra \la E, d |
\eqno(4.4)
$$
where $ | E, d \ra $ are the energy eigenstates with
degeneracy label $d$.

The standard coherent states (see Ref.[\cite{Gar}], for example)
are denoted $ | \p, \x \ra $ and they
have the important property that they are preserved in form under
unitary evolution,
$$
e^{ - i H_0 t } | \p, \x \ra = | \p_t, \x_t \ra
\eqno(4.5)
$$
where $ \p_t, \x_t $ are the classical solutions matching
$\p, \x $ at $t=0$, hence they are strongly peaked about
the classical path. We are interested in finding a set of states
which are analogues of these states for the timeless case. That is,
they should be eigenstates of $H_0$, and should be peaked about the
classical paths of given fixed energy in
phase space. It is not difficult to see that a set of states doing
the job are,
$$
\eqalignno{
| \phi_{\p \x} \ra &=  \delta (H_0 - E ) | \p, \x \ra
\cr
&=  \int_0^{2 \pi} { dt \over 2 \pi } \ e^{ - i (H_0 - E ) t }
\ | \p , \x \ra
\cr
&=   \int_0^{2 \pi} { dt \over 2 \pi } \ e^{i E t }\ | \p_t, \x_t \ra
&(4.6) \cr}
$$
These states are not in fact normalized to unity but we shall
see that it is useful to work with them as they are.
Since the states $ | \p_t, \x_t \ra $ are concentrated at a phase
space point for each $t$, clearly integrating $t$ over a whole period
produces a state which is concentrated along the entire classical
trajectory. Each state is labeled by a fiducial phase
space point $\p, \x$ which
determines the classical trajectory the state is peaked about.
Under evolution of the fiducial point $\p, \x$ to another point,
$\p_s, \x_s$, say, along the same classical trajectory, the
state changes by a phase,
$$
| \phi_{\p \x} \ra \ \ria \  | \phi_{\p_s \x_s} \ra =
e^{ i E s } | \phi_{\p \x} \ra
\eqno(4.7)
$$
as may be seen from (4.5), (4.6). We will refer to
these states as {\it timeless coherent states}. Their properties are in fact
very similar to the usual coherent states.

Note that the coherent states $ | \p, \x \ra $ are
in fact already {\it approximate} eigenstates of $H_0$, with
eigenvalue $ \half (\p^2 + \x^2 ) $, as long
as $|\p|, |\x| $ are much larger than the coherent state
quantum fluctuations. Given $E$, it therefore seems reasonable
to choose the values $\p, \x$ in the fiducial coherent state
so that $ E = \half (\p^2 + \x^2 ) $, when constructing
the timeless states (4.6).

Two timeless coherent states of different energy are exactly
orthogonal. The more interesting case is that in which
they have the same energy, and then they are approximately
orthogonal if they correspond to sufficiently distinct classical
solutions. This is because we have
$$
\eqalignno{
\la \phi_{\p' \x'} | \phi_{\p \x} \ra
& =   \la \p', \x' | \delta (H_0 - E ) | \p, \x \ra
\cr
& =   \int_0^{2 \pi} { dt \over 2 \pi}
\ e^{ i E t }
\la \p', \x' | \p_t, \x_t \ra
&(4.8) \cr }
$$
From the properties of the standard coherent states
we know that
$$
| \la \p', \x' | \p_t, \x_t \ra | \le 1
\eqno(4.9)
$$
with equality
if and only if $ \p' = \p_t $ and $\x' = \x_t $. Moreover,
the overlap of two coherent states is exponentially small
if they are centered around phase space points that are
sufficiently far apart. It follows that
if $\p', \x'$ does not lie on, or close to, the trajectory
$\p_t, \x_t$, the overlap $ \la \p', \x' | \p_t, \x_t \ra $
will always exponentially small for all $t$. The integral over
$t$ in (4.8) will then give a result that is much smaller than
the case in which $\p', \x'$ does lie on, or close to, the trajectory
$\p_t, \x_t$ (because in the latter case the overlap
$ \la \p', \x' | \p_t, \x_t \ra $ becomes close to unity
for some value of $t$).
The timeless coherent states are therefore approximately
orthogonal for sufficiently distinct classical trajectories.

The standard completeness relation for the coherent states is
$$
\int {d^N \p \ d^N \x \over (2 \pi )^N }
\ | {\p\x} \ra \la {\p \x} | = 1
\eqno(4.10)
$$
Multiplying both sides by $\delta (H_0 - E)$ from the left and right,
and using (4.6), we get
$$
\int {d^N \p \ d^N \x \over (2 \pi )^N }
\  |\phi_{\p\x} \ra \la \phi_{\p \x} |  =  \delta (H_0 - E)
\eqno(4.11)
$$
Since $ \delta (H_0 - E ) | \psi \ra = | \psi \ra $ on any solution
to the eigenvalue equation $(H_0 - E) | \psi \ra = 0 $, this is
as good as a completeness relation on the set of solutions
to the eigenvalue equation (which is
all we are interested in). We may therefore write any solution
$|\psi \ra$ as a superposition of timeless coherent states,
$$
| \psi \ra = \int {d^N \p \ d^N \x \over (2 \pi )^N}
\  | \phi_{\p \x } \ra \la \phi_{\p \x} |\psi \ra
\eqno(4.12)
$$
It is then tempting to interpret
$  \la \phi_{\p \x }| \psi \ra $ as the amplitude
that a system in the state $ | \psi \ra $ will be found on the classical
trajectory corresponding to the timeless coherent state
$ | \phi_{\p \x} \ra $.
Again using the fact that $ \delta (H_0 - E ) | \psi \ra = | \psi \ra $
it is easy to see that this is in fact the same as $ \la \p,
\x | \psi \ra  $, which is the amplitude for finding the
system at the phase space point $ \p, \x $ labeling the
trajectory. The probability is then simply $ | \la \p,
\x | \psi \ra |^2 $. This is a simple and intuitively
appealing result: the classical trajectory is completely
fixed by its initial values $\x, \p$, hence we expect
that the probability for being found on a certain classical
trajectory is the same as the probability for being found
at the initial phase space point that labels it.

This interpretation is put forward with a small note of caution,
however, since the sum over $\p, \x$ in Eq.(4.12) is not
only over states which are only approximately orthogonal
(as with the usual coherent states), but, because of the
property (4.7), includes some redundancy in the summation.
In particular, $| \phi_{\p \x } \ra \la \phi_{\p \x} |$
is invariant along the classical phase space trajectory
of its fiducial point. Since this only produces some sort of
constant factor, it may not make any difference, and indeed,
the above interpretation appears to produce sensible results.
Still, it would be desirable to include, if possible, some sort
of ``gauge fixing'' which factors out this redundancy.
This will be explored elsewhere.

Given all of this background, we may now reconsider the detector
amplitude (2.25). We
expand the initial state $ | \psi \ra $ in the timeless coherent
states, as in Eq.(4.12). When $f_1$ operates on the state, it has the effect
of restricting the sum over timeless coherent states to only those
that pass through the region $\Delta_1$, but otherwise leaves the
timeless coherent states preserved in form (as long as the region
$\Delta_1$ is larger than the width of the coherent states).
Next, to consider
the operation of $G$, we write it as
$$
G = i \int_0^{\infty} d \tau \ e^{ - i \tau (H - i \e ) }
\eqno(4.13)
$$
where $H = H_0 - E $. When the timeless coherent states
are operated on with the exponential factor in (4.13)
it has essentially no effect, since $ H | \phi_{\p \x} \ra = 0 $.
(The time integration leads to a constant factor, which may, however,
need regulating along the lines of (1.6)). Operating with $f_2$
further restricts the sum to those states that pass through both
$\Delta_1$ and $\Delta_2$, and then subsequent operation $G$
again does essentially nothing. And likewise to
the end of the chain. We
therefore find that the detector amplitude (2.25) consists of a
superposition of only those timeless coherent states that pass
through all the detector regions $ \Delta_1 $, $ \Delta_2 \cdots
\Delta_n$. This could be quite a large sum of states if the
regions $\Delta_k$ are large. However, it will consist of
essentially just one timeless coherent state if the detector
regions lie along a classical trajectory and if their
size is just bigger than the spatial width of the wave packet.
The detection amplitude along this classical trajectory
is then equal to $ \la \phi_{\p \x} | \psi \ra $, in
agreement with the analysis based on Eq.(4.12).

Rovelli has written down a coherent state of the type
considered here
[\cite{Rov1}] in the context of a very similar model, although its
properties were not explored and exploited as they are here.
Klauder, in his approach to the quantization of constrained
systems using coherent states, considered the projection of the
standard coherent states onto the contraint subspace, hence in essence
wrote down states of the form (4.6) [\cite{Kla}]. He did not,
however, consider their use as an interpretational tool.
Wave packet solutions to the Wheeler-DeWitt equation which
aproximately track the classical trajectories for more interesting
cosmological models have been considered in
Refs.[\cite{Kie,KaNa}].

\head{\bf 5. Exact Solution to the Detector Dynamics}

In this Section we discuss an alternative method of
solution of the eigenvalue equation (2.2). As we shall see,
it does not in fact give a very elegant representation
of the Mott solution, which is why it was not used above.
If, however, one is permitted to use different boundary conditions,
as may be reasonable in quantum cosmology, then it provides
an alternative possible solution to the detector dynamics.

We consider first the case of a single detector.
The key observation is that solution to the eigenvalue equation
(2.2) may be generated using either of the expressions,
$$
\eqalignno{
| \Psi \ra &= \delta (H - E ) | \phi \ra
&(5.1)
\cr
&= {1 \over 2 \pi}
\int_{-\infty}^{\infty} d \tau \ e^{ -i \tau ( H_0 + \l H_{int} - E ) }
| \phi \ra
&(5.2) \cr}
$$
Here $| \phi \ra $ is an arbitrary fiducial state
in the joint system-detector Hilbert space. It is ambiguous
up to the addition of a term of the form $(H-E) | \phi' \ra $.
For the moment we keep $| \phi \ra $ general and take
$$
| \phi \ra = | \chi_0 \ra | 0 \ra + | \chi_1 \ra | 1 \ra
\eqno(5.3)
$$
To evaluate (5.1) or (5.2) we introduce the eigenstates of $a + a^{\dag}$,
which are
$$
| \pm \ra = { 1 \over \sqrt{2} } \left( | 0 \ra \pm | 1 \ra \right)
\eqno(5.4)
$$
The expression (5.1) is then readily evaluated with
the result
$$
| \Psi \ra = | \psi_{nd} \ra | 0 \ra + | \psi_d \ra | 1 \ra
\eqno(5.5)
$$
where,
$$
\eqalignno{
| \psi_{nd} \ra &=
\half \left( \delta ( h + \l f_1 ) +\delta ( h - \l f_1 ) \right) | \chi_0 \ra
+ \half \left( \delta ( h + \l f_1 ) - \delta ( h - \l f_1 ) \right) | \chi_1 \ra
\quad
&(5.6) \cr
| \psi_{d} \ra &=
\half \left( \delta ( h + \l f_1) -\delta ( h - \l f_1) \right) | \chi_0 \ra
+ \half \left( \delta ( h + \l f_1 ) + \delta ( h - \l f_1 ) \right) | \chi_1 \ra
\quad
&(5.7) \cr}
$$
(recalling that $h = H - E $).
These expressions are perhaps more easily appreciated using a path
integral respresentation:
$$
\eqalignno{ \psi_{nd} (\x_f )
=& \int_{-\infty}^{\infty} d \tau
\ e^{i \tau E} \int {\cal D} \x (t) \exp \left( i S [\x (t) ] \right)
\cr
& \times \left[ \cos \left( \lambda \int_0^{\tau} dt f_1(\x (t) )
\right)  \chi_0 (\x_0 )
+i  \sin \left( \lambda \int_0^{\tau} dt f_1(\x (t) ) \right)
\chi_1 (\x_0 )
\right]
&(5.8) \cr
\psi_{d} (\x_f )
=&
\int_{-\infty}^{\infty} d \tau \ e^{i \tau E} \int {\cal D} \x (t)
\exp \left( i S [\x (t) ] \right)
\cr
& \times \left[ i \sin  \left( \lambda \int_0^{\tau} dt f_1(\x (t) )
\right)  \chi_0 (\x_0 )
+ \cos \left( \lambda \int_0^{\tau} dt f_1(\x (t) )
\right)  \chi_1 (\x_0 )
\right]
&(5.9) \cr}
$$
Eqs.(5.8), (5.9) represent the exact solution to Eq.(2.2)
with one detector in place.

Turn now to the question of the fiducial state. The condition
(2.5), suggests that we should take
$$
| \phi \ra = | \chi_0 \ra | 0 \ra
\eqno(5.10)
$$
where $ | \psi \ra = \delta (H_0 - E ) | \chi_0 \ra $,
and hence that $ | \chi_1 \ra = 0 $.
In fact, in the induced inner product scheme, we may
take $ | \chi_0 \ra = | \psi \ra $, since effectively
$ \delta (H_0 - E )^2 = \delta (H_0 - E ) $. This therefore
yields a path integral expression for the amplitude for
detection and has the property that the factoring condition
(2.5) is satisfied when $\l = 0$.
So let us first explore the properties of (5.8), (5.9)
with $\chi_1 = 0 $.

The nature of the sums over paths $\x (t)$ in Eq(5.8) and (5.9) is
governed by the quantity
$$
\tau_d =\int_0^{\tau} dt f_1(\x (t) )
\eqno(5.11)
$$
appearing in the sine and cosine factors. With $f_1$ normalized to
be dimensionless, this quantity has the dimensions of time, and
is essentially the time spent by the path $\x(t)$ in the region
$\Delta_1$ around the detector. (This is not of course a
physically measurable time. The paths $\x(t)$ in the path integral
have a well-defined notion of time, but after summing over the
total time duration of each path $\tau$ the final result is
time-independent). Now, in the amplitude for no detection, (5.8),
the factor $\cos(\lambda \tau_d)$ is $1$ for $\tau_d =0$, and
decreasing for $\tau_d$ increasing from zero. In the path
integral, it therefore has the effect of suppressing paths that
pass close to the detector, and favours paths that stay away from
it. Similarly, in the detection amplitude, (5.9), the factor
$\sin(\lambda \tau_d)$ is zero for paths that spend no time near
the detector, and non-zero for paths that enter the detector
region. It therefore enhances the amplitude for paths entering the
detector region. The sine and cosine factors therefore, in a very
crude way, enforce restrictions  on the paths corresponding to
entering or not entering the detector region, as one would expect.
We can also see, however, that these factors only do their job
well if $\lambda \tau_d $ is somewhat smaller than $1$, indicating
that the detector model is only physically sensible in the
perturbative regime about $\l = 0$, as expected.

It is now important to check the agreement between the exact
result above and the perturbative result of
Section 2. It is not difficult
to show that the exact solution (5.9) with $\chi_1 = 0$
does not in fact agree with the perturbative solution (2.19).
It differs by the presence of homogeneous solutions in the
small $\l$ limit of (5.9). We will not go into details
but it may be shown using (5.7) and the identity,
$$
\delta ( h + \l f) = \delta ( h ) + i \l
\left( G f G - G^{\dag} f G^{\dag} \right) + O (\l^2)
\eqno(5.12)
$$
(which is proved using the exponential representation (5.2)).
Recall that in the perturbative
solution the homogeneous solutions were removed at each
order in perturbation theory essentially by inspection.
Since no corresponding condition has been imposed here,
it is not surprising that these
spurious solutions crop up. We can, however,
get agreement if we choose
$| \chi_1 \ra = i \l f G^{\dag} | \chi_0 \ra $
in Eq.(5.9) (again proved using (5.12),
although there is no obvious independent
reason for making this choice. Furthermore, the presence
of the Green function $G^{\dag}$ in the fiducial state
rather destroys the elegance of the path integral
representation compared to the case $\chi_1 = 0$.

On the other hand, although the Mott solution is not
readily obtained, the removal of the homogeneous
solutions in the perturbative solution is a subtle matter
of boundary conditions, above and beyond the basic factoring
condition (5.10). As discussed in Section 3,
in the case of quantum cosmology one
is not obviously obliged to take the same conditions,
and indeed one can argue that, beyond (5.10),
the boundary conditions are up for grabs. Indeed, one
could effectively choose boundary conditions by
{\it proposing} that the solution is given by the
formula (5.1) with the factoring condition (5.10).
As one can see from (5.12), this proposal again produces
solutions with a time symmetric flavour to them
({\it i.e.}, with an equal number of $G$'s and $G^{\dag}$'s
in the solution), a feature which persists to the case
of more than one detector. Not surprisingly, this choice does
in fact produce, in essence, the Hartle amplitude (3.8).

We have, therefore, in this Section produced another candidate
expression for the detection amplitude, which is arguably the more
appropriate one for quantum cosmology. Furthermore, the means of
generating it, Eqs.(5.1), (5.2), are readily generalizable to more
complicated situations.

\head{\bf 6. An Improved Detector Model}

We now briefly consider a more elaborate detector model that
consists of a harmonic oscillator instead of the simple two state
system. So we take, in the case of a single detector,
$$
H_d = \omega a^{\dag} a, \quad H_{int} = f (\x) (a + a^{\dag})
\eqno(6.1)
$$
The energy eigenstate of the total Hamiltonian is again
calculated using (5.2), with the factored fiducial state (5.10),
where $ |0\ra $ is the harmonic oscillator ground state.
In terms of a path integral, denoting the harmonic oscillator
coordinates by $q$,
$$
\eqalignno{
\Psi (\x_f, q_f ) = & \int_{-\infty}^{\infty} d \tau
\ e^{ i E \tau} \ \int {\cal D} \x(t) {\cal D} q (t)
\cr
& \times \ \exp \left( i S_0 [\x(t)] + i S_d [q(t)] + i \lambda S_{int}
[q(t), \x (t)  ] \right) \ \chi_0 (\x_0 ) u_0 (q_0 )
&(6.2) \cr }
$$
where $u_0 (q) = \la q | 0 \ra $. The integral over $q$ is conveniently
rewritten as
$$
\eqalignno{
\int {\cal D} q(t) & \exp \left( i S_d [q(t)] + i \lambda
S_{int} [ q(t), \x (t) ] \right) \ u_0 (q_0)
\cr & \quad \quad \quad
= \la q_f | T \exp \left( - i \tau ( H_d + \l H_{int} ) \right) | 0 \ra
&(6.3) \cr }
$$
where $T$ denotes time ordering. The right-hand side of (6.3)
is just the unitary evolution
of the vacuum state for a driven harmonic oscillator.
Using the properties of coherent states [\cite{Gar}], (6.3) is equal to
$ \la q_f | z(\tau ) \ra $, where $ | z \ra $ is a standard coherent state
and
$$
z(\tau) = -i \l \int_0^{\tau} dt \ e^{i \omega (\tau - t )} \ f ( \x (t) )
\eqno(6.4)
$$
To find the amplitudes for detection and no detection we expand
the total state (6.2) in terms of the eigenstates of the harmonic
oscillator. Since all states other than the ground state correspond
to detection, there is no single amplitude corresponding to
detection (although there is a probability). It is therefore easier
to look at the amplitude for no detection, which is obtained
by overlapping (6.2) with the ground state $u_0 (q_f)$, yielding,
$$
\psi_{nd} (\x_f ) =  \int_{-\infty}^{\infty} d \tau
\ e^{ i E \tau} \ \int {\cal D} \x(t) \ \la 0 | z (\tau) \ra
\ \exp \left( i S_0 [\x(t)] \right)
\chi_0 (\x_0 )
\eqno(6.5)
$$
From the properties of coherent states [\cite{Gar}], we have
$$
\la 0 | z (\tau ) \ra = \exp \left( - \half | z (\tau) |^2 \right)
\eqno(6.6)
$$
The probability for no detection is $ \la \psi_{nd} | \psi_{nd} \ra$
and the probability for detection is simply
$ 1 - \la \psi_{nd} | \psi_{nd} \ra $.

The result (6.5) clearly has the desired properties. For paths $\x(t)$
which never enter the detection region, $ z(\tau) = 0 $ and the
path integral is unaffected. Paths that enter the region, on
the other hand, generally have $z ( \tau ) \ne 0 $, and they
are exponentially suppressed. This is therefore a much
improved detector model in comparison to Eq.(5.8). Its validity is
not restricted to the perturbative regime. The generalization
to many detectors is trivial and essentially the same
result concerning peaking about classical trajectories
is the obtained.

Of course, this detector is still not fully satisfactory because it can
happen that $z (\tau) = 0 $ even for paths that enter the region,
because of the oscillatory nature of $z (\tau) $, hence we
again encounter the issue of detector recurrences. Again this
will be avoided if the time the trajectory spends in the detector
region is short (less than $\omega^{-1}$).

A more challenging detector improvement avoiding the recurrence
problem would be to construct
one that, were it used in standard unitary quantum mechanics,
would be irreversible, {\it i.e.}, involves an essentially
infinite number of degrees of freedom. Such a detector
was introduced in the related context of measuring arrival
times in Ref.[\cite{Hal5}] and it would be interesting to incorporate
it into the situation considered here.

\head{\bf 7. Summary}

The aim of this paper was to give substance to the appealing
intuitive notion that solutions to the Wheeler-DeWitt equation
(1.1) correspond to entire histories of the universe with
time emerging as a parameter along each trajectory.
The concrete technical results -- the detection amplitude
and the introduction of a set of timeless coherent states
-- are compatible with this notion. There are, however, many subtle
aspects to this notion [\cite{Bar3}], and we do not claim to have
an exhaustive demonstration of the emergence of trajectories
from the Wheeler-DeWitt equation.

\head{\bf Acknowledgements}

I am very grateful to Julian Barbour, Jim Hartle, John Klauder
and Dieter Zeh for their comments on the first draft of this
paper.

\references

\def\pr{{\sl Phys. Rev.\ }}
\def\prl{{\sl Phys. Rev. Lett.\ }}

\def\jmp{{\sl J. Math. Phys.\ }}

\def\np{{\sl Nucl. Phys.\ }}

\def\annp{{\sl Ann. Phys. (N.Y.)\ }}
\def\cqg{{\sl Class. Quant. Grav.\ }}

\refis{Ash} For a nice review see, C.Rovelli, gr-qc/9710008,
{\it Loop quantum gravity}.

\refis{asy} See, for example, the collection of articles in {\it
Physical Origins of Time Asymmetry}, edited by J.J.Halliwell,
J.Perez-Mercader and W.Zurek (Cambridge University Press,
Cambridge, 1994). A recent discussion is K.Ridderbos, {\sl Studies
in History and Philosophy of Modern Physics} {\bf 30}, 41(1999).

\refis{Bar1} J.Barbour, \cqg {\bf 11}, 2853 (1994).

\refis{Bar2} J.Barbour, \cqg {\bf 11}, 2875 (1994).

\refis{Bar3} J.Barbour, {\it The End of Time: The Next Revolution
in our Understanding of the Universe} (Weidenfeld and Nicholson,
1999).

\refis{Bel} J.S.Bell, {\it Speakable and Unspeakable in Quantum Mechanics}
(Cambridge University Press, Cambridge, 1987).

\refis{Bro} See also, A.A.Broyles, \pr {A48}, 1055 (1993), and
M.Castagnino and R.Laura, gr-qc/0006012, for further discussions
of the Mott calculation.

\refis{BuIs} J.Butterfield and C.J.Isham, gr-qc/9901024.

\refis{CoHe} K.Hepp, {\sl Helv.Phys.Acta} {\bf 45}, 237 (1972).

\refis{DeW} B.DeWitt, in {\it Gravitation: An Introduction to
Current Research}, edited by L.Witten (John WIley and Sons, New
York, 1962).

\refis{DH} M.Gell-Mann and J.B.Hartle, in {\it Complexity,
Entropy and the Physics of Information, SFI Studies in the
Sciences of Complexity}, Vol. VIII, W. Zurek (ed.) (Addison
Wesley, Reading, 1990); and in {\it Proceedings of the Third
International Symposium on the Foundations of Quantum Mechanics in
the Light of New Technology}, S. Kobayashi, H. Ezawa, Y. Murayama
and S. Nomura (eds.) (Physical Society of Japan, Tokyo, 1990);
{\sl Phys.Rev.} {\bf D47}, 3345 (1993); R.B.Griffiths, {\sl
J.Stat.Phys.} {\bf 36}, 219 (1984); {\sl Phys.Rev.Lett.} {\bf 70},
2201 (1993); {\sl Am.J.Phys.} {\bf 55}, 11 (1987); R.Omn\`es, {\sl
J.Stat.Phys.} {\bf 53}, 893 (1988); {\bf 53}, 933 (1988); {\bf
53}, 957 (1988); {\bf 57}, 357 (1989); {\bf 62}, 841 (1991); {\sl
Ann.Phys.} {\bf 201}, 354 (1990); {\sl Rev.Mod.Phys.} {\bf 64},
339 (1992); J.B.Hartle, in {\it Quantum Cosmology and Baby
Universes}, S. Coleman, J. Hartle, T. Piran and S. Weinberg (eds.)
(World Scientific, Singapore, 1991); J.J.Halliwell, in {\it
Fundamental Problems in Quantum Theory}, edited by D.Greenberger
and A.Zeilinger, Annals of the New York Academy of Sciences, Vol
775, 726 (1994). For further developments in the decoherent
histories approach, particularly adpated to the problem of
spacetime coarse grainings, see C. Isham, \jmp {\bf 23}, 2157
(1994);
C. Isham and N. Linden, \jmp {\bf 35}, 5452 (1994); {\bf 36}, 5392
(1995).

\refis{Gar} C.W.Gardiner, {\it Quantum Noise} (Springer-Verlag,
Berlin, 1991).

\refis{Gol} H.Goldstein, {\it Classical Mechanics}
(Addison-Wesley, Reading MA, 1980).

\refis{Hal1} J.J.Halliwell, in, {\it Proceedings of the 13th
International Conference on General Relativity and Gravitation},
edited by R.J.Gleiser, C.N.Kozameh, O.M.Moreschi
(IOP Publishers, Bristol,1992). (Also available as
the e-print gr-qc/9208001).

\refis{Hal2} J.J.Halliwell, \pr {\bf D48}, 4785 (1993).

\refis{Hal3} J.J.Halliwell, \pr {\bf D38}, 2468 (1988).

\refis{Hal4} This detector model was used in a simple
non-relativistic context by J.J.Halliwell, \pr {\bf D60}, 105031
(1999).
Some subsequent developments of the Coleman-Hepp model are
H.Nakazato and S.Pascazio, \prl {\bf 70}, 1 (1993); \pr {\bf A48},
1066 (1993); R.Blasi, S.Pascazio, S.Takagi, \pr {\bf A250}, 230 (1998).

\refis{Hal5} J.J.Halliwell, {\sl Prog.Theor.Phys.} {\bf 102}, 707 (1999).

\refis{HaOr} J.J.Halliwell and M.E.Ortiz, {\sl Phys.Rev.} {\bf D48}, 748 (1993).

\refis{HaTh} J.J.Halliwell and J.Thorwart, in preparation.
See also the related works, J.Whelan, \pr {\bf D50}, 6344 (1994);
J.B.Hartle and D.Marolf,  \pr {\bf D56}, 6247 (1997).


\refis{Har2} J.B.Hartle, \pr {\bf D38}, 2985 (1988).

\refis{Har3} J.B.Hartle, in {\it Proceedings of the 1992 Les Houches
School, Gravity and its Quantizations}, edited by B.Julia and
J.Zinn-Justin (Elsevier Science B.V. 1995)

\refis{HaHa} J.B.Hartle and S.W.Hawking, \pr {\bf 28}, 2960
(1983).

\refis{Haw} S.W.Hawking, \pr {\bf D32}, 2489 (1985).

\refis{Ish} C.J.Isham, gr-qc/9210011.

\refis{KaNa} Y.Kazama and R.Nakayama, \pr {\bf 32}, 2500 (1985).

\refis{Kie} C.Kiefer, \pr {\bf D38}, 1761 (1988).

\refis{Kla} J.Klauder, Ann. Phys. (NY) {\bf 254}, 419 (1997),
(quant-ph/9604033) ; Nucl.Phys. {\bf B547},397, 1999,
(hep-th/9901010); hep-th/0003297.

\refis{Kuc} K.Kuchar, in {\it Conceptual Problems of Quantum
Gravity}, edited by A.Ashtekar and J.Stachel (Boston, Birkhauser,
1991); and in {\it Proceedings of the 4th Canadian Conference on
General Relativty and Relativistic Astrophysics}, edited by
G.Kunstatter, D.E.Vincent and J.G.Williams (World Scientific, New
Jersey, 1992). See also the e-print gr-qc/9304012,
{\it Canonical quantum gravity}.

\refis{Mar1} D.Marolf, \cqg {\bf 12}, 1199 (1995).

\refis{Mar2} D.Marolf, \pr {\bf D53}, 6979(1996);
\cqg {\bf 12}, 2469 (1995);
\cqg{\bf 12}, 1441 (1995).

\refis{Mot} N.F.Mott, {\sl Proc.Roy.Soc} {\bf A124}, 375 (1929),
reprinted in {\it Quantum Theory and Measurement}, edited by
J.Wheeler and W.Zurek (Princeton University Press, Princeton, New
Jersey, 1983).

\refis{PDX} A.Auerbach and S.Kivelson, \np {\bf B257}, 799 (1985).

\refis{Rie} A.Ashtekar, J.Lewandowski, D.Marolf, J.Mourao and
T.Thiemann, {\sl J.Math.Phys.} {\bf 36}, 6456 (1995);
A.Higuchi, \cqg {\bf 8}, 1983 (1991).
D.Giulini and D.Marolf, \cqg {\bf 16}, 2489 (1999);
\cqg {\bf 16}, 2479 (1999).
F.Embacher, {\sl Hadronic J.} {\bf 21}, 337 (1998);
N.Landsmann, {\sl J.Geom.Phys.} {\bf 15}, 285 (1995).

\refis{Rov1} C.Rovelli, \pr {\bf 42}, 2638 (1990).

\refis{Rov2} C.Rovelli, \pr {\bf 43}, 442 (1991).

\refis{Rov3} C.Rovelli, \cqg {\bf 8}, 297 (1991);
{\bf 8}, 317 (1991).

\refis{Sch} L.Schulman, {\it Techniques and Applications of Path
Integrals} (Wiley, New York, 1981).

\refis{Tei} C.Teitelboim,
\pr {\bf D25}, 3159 (1983); {\bf 28}, 297 (1983);
{\bf 28}, 310 (1983).

\refis{Time}
Y.Aharanov and D.Bohm, \pr {\bf 122}, 1649 (1961);
Y.Aharanov, J.Oppenheim, S.Popescu, B.Reznik and
W.Unruh, quant-ph/9709031 (1997);
G.R.Allcock, \annp {\bf 53}, 253 (1969);
{\bf 53}, 286 (1969); {\bf 53}, 311 (1969);
Ph.Blanchard and A.Jadczyk,
{\sl Helv.Phys.Acta.} {\bf 69}, 613 (1996);
I.Bloch and D.A.Burba, \pr {\bf 10}, 3206 (1974);
V.Delgado, preprint quant-ph/9709037 (1997);
R.Giannitrapani, preprint quant-ph/9611015 (1998);
N.Grot, C.Rovelli and R.S.Tate, \pr {\bf A54}, 46 (1996);
E.Gurjoy and D.Coon, {\sl Superlattices and
Microsctructures} {\bf 5}, 305 (1989);
A.S.Holevo, {\it Probabilistic and Statistical Aspects
of Quantum Theory} (North Holland, Amsterdam, 1982), pages 130--197;
A.Jadcyk, {\sl Prog.Theor.Phys.} {\bf 93}, 631 (1995);
D.H.Kobe and V.C.Aguilera--Navarro,  \pr {\bf A50}, 933 (1994);
N.Kumar, {\sl Pramana J.Phys.} {\bf 25}, 363 (1985);
J.Le\'on, preprint quant-ph/9608013 (1996);
D.Marolf, \pr {\bf A50}, 939 (1994);
L.Mandelstamm and I.Tamm, {\sl J.Phys.} {\bf 9}, 249 (1945);
J.G.Muga, S.Brouard and D.Mac\'ias, \annp {\bf 240}, 351 (1995);
J.G.Muga, J.P.Palao and C.R.Leavens, preprint quant-ph/9803087
(1987);
J.G.Muga, R.Sala and J.P.Palao, preprint quant-ph/9801043,
{\sl Superlattices and Microstructures} {\bf 23} 833 (1998);
C.Piron, in {\it
Interpretation and Foundations of Quantum Theory}, edited by
H.Newmann (Bibliographisches Institute, Mannheim, 1979);
M.Toller, preprint quant-ph/9805030 (1998);
H.Salecker and E.P.Wigner, \pr {\bf 109}, 571 (1958);
F.T.Smith, \pr {\bf 118}, 349 (1960);
E.P.Wigner, \pr {\bf 98}, 145 (1955).

\refis{YaT} N.Yamada and S.Takagi, {\sl Prog.Theor.Phys.}
{\bf 85}, 985 (1991); {\bf 86}, 599 (1991); {\bf 87}, 77 (1992);
N. Yamada, {\sl Sci. Rep. T\^ohoku Uni., Series 8}, {\bf 12}, 177
(1992); \pr {\bf A54}, 182 (1996); J.J.Halliwell and E.Zafiris,
{\sl Phys.Rev.} {\bf D57}, 3351-3364 (1998);
J.B.Hartle, {\sl Phys.Rev.} {\bf D44}, 3173 (1991);
R.J.Micanek and J.B.Hartle, {\sl Phys.Rev.} {\bf A54},
3795 (1996).

\refis{Zeh} H.D.Zeh, {\it The Physical Basis of the Direction of
Time}, third edition (Springer-Verlag, 1999) (and the associated
webpage www.time-direction.de); {\sl Phys.Lett.} {\bf A116}, 9
(1986); {\bf A126}, 311 (1988); C.Kiefer and H.Zeh, {\sl
Phys.Rev.} {\bf D51}, 4145 (1995).

\endreferences

\end